
\input phyzzx

\REF\MSW{G. Mandal, A. M. Sengupta and S. R. Wadia, Mod. Phys. Lett. A6
(1991) 1685.}
\REF\WITTEN{E. Witten, Phys. Rev. D44 (1991)}
\REF\EFR{S. Elitzur, A. Forge and E. Rabinovichi, Hebrew Univ. Preprint
RI-143/90.}
\REF\RSS{M. Rocek, K. Schoutens and A. Sevrin, IAS Preprint,
IASSNS-HEP-91/14.}
\REF\OTHERS{M. Bershadsky and D. Kutasov, Princeton Preprint, PUPT-1261,
HUTP-91/A024., E. Martinec and S. Shatashvili, Chicago Preprint,
EFI-91-22 }
\REF\DVV{R. Dijkgraaf, E. Verlinde and H. Verlinde, Princeton Preprint,
PUPT-1252, IASSNS-HEP-91/22}
\REF\GIV{A. Giveon, Berkeley Preprint, LBL-30671, 1991}
\REF \BIPZ{E. Br\'ezin, C. Itzykson,
G. Parisi and J.B. Zuber, Comm. Math. Phys. 59, 35 (1978); E.Brezin,
V.A.Kazakov and Al.B.Zamolodchikov, Nucl. Phys. B338(1990)673; D.Gross and
N.Miljkovic, Nucl. Phys. B238(1990) 217; P.Ginsparg
and J.Zinn-Justin, Phys. Lett. 240B(1990)333; G.Parisi, Europhys. Lett.
11(1990)595; S. R.Das, A.Dhar, A.M.Sengupta and S. R. Wadia, Mod.Phys. Lett. A5
(1990) 891.}

\REF\POLTACH{J. Polchinski,  Nucl. Phys. B346 (1990)253.}
\REF\DJ{S. Das and A. Jevicki, Mod. Phys. Lett. A5(1990) 1639.}
\REF\SW{A. M. Sengupta and S. R. Wadia, Int. J. Mod. Phys. A6 (1991) 1961.}
\REF\GK{D. Gross and I. Klebanov, Nucl. Phys. B352 (1990) 671. }
\REF\MSWTACH{G. Mandal, A. M. Sengupta and S. R. Wadia,
Mod. Phys. Lett. A6 (1991) 1465.}
\REF\M{G. Moore, Yale and Rutgers preprint, YCTP-P8-91, RU-91-12.}
\REF\POLCHINSKI{J.Polchinski, Texas Preprint, UTTG-06-91.}
\REF\DJR{K. Demeterfi, A. Jevicki and J. Rodrigues, Brown Preprint,
BROWN-HET-795, 1991}
\REF\POLYAKOV{A.Polyakov, Mod. Phys. Lett. A6 (1991) 635.}
\REF\KD{D.Kutasov and P.Difrancesco, Princeton preprint, PUPT-1237,1991.}
\REF\GROSS{D.Gross and I.Klebanov, Princeton preprint, PUPT-1241,
February 1991.}
\REF\GKCIRC{D. Gross and I. Klebanov, Nucl. Phys. B344 (1990) 475.}
\REF\DN{J. Distler and P. Nelson, Princeton Preprint, UPR-0462T, PUPT-1262,
1991.}
\REF\KT{Ya. Kogan, JETP Lett. 45 (1987) 709; B. Sathyapalan, Phys. Rev.
D35 (1987) 3277.}
\REF\GPY{D. Gross, M. Perry and L. Yaffe, Phys. Rev. D25 (1982) 330. }

\hfill{TIFR-TH-91/45}\break
\indent\hfill{November, 1991}\break
\date{}
\titlepage
\title {INSTABILITIES IN THE GRAVITATIONAL BACKGROUND AND STRING THEORY}
\author{Anirvan M. Sengupta \foot{e-mail : anirvan@tifrvax.bitnet}}
\address {Tata Institute of Fundamental Research,
Homi Bhabha Road, Bombay
400 005, INDIA}

\abstract { We indicate the tentative source of instability  in the
two-dimensional black hole background. There are relevant
operators among the
 tachyon  and the higher level vertex operators in the conformal
field theory. Connection of this instability with Hawking
radiation is not obvious. The situation is somewhat analogous to
fields in the background of a negative mass Euclidean
Schwarzschild solution (in four dimensions). Speculation is made
about decay of the Minkowski black hole
into finite temperature flat space.}


\endpage

\chapter{ Introduction}
Recently there has been some interest in the classical
gravitational background consistent with string theory in a
two-dimensional target space[\MSW,\WITTEN,\EFR,\RSS,\OTHERS,\DVV,
\GIV]. If one looks at the
graviton-dilaton sector, one finds a black hole solution (or
one of its relatives). It has been conjectured [\WITTEN] that
the Hawking radiation would transform this black hole to an extreme
Reissner-Nordstrom-like solution, which would then
represent the $c = 1$ matrix model [\BIPZ,\GKCIRC,
\POLTACH,\DJ,\SW,\GK] and the particle absorption
and re-emission by this terminal black hole would be the
tachyon scattering near the wall [\MSWTACH,\M,\POLCHINSKI,\DJR
\POLYAKOV,\KD,\GK,\GROSS], in the matrix model picture.

In any case, it is of some importance to understand how the
various classical solutions in 2-d string theory are related
to each other. The $c = 1$ matrix model seems to be
a solution with a nontrivial dilaton and tachyon background but
with a flat metric [\POLTACH,\DJ,\SW,\GK]. The black hole solution
has nontrivial dilaton-graviton background in which we can introduce tachyon
perturbations [\MSW,\WITTEN,\EFR,\RSS].
We suggest that the second solution is an
unstable stationary point. This instability  would make it go over to a
different spacetime geometry, may be a flat spacetime with one compact
direction.

However, it is not clear that this instability has something to
do with Hawking radiation. Usually, in static black hole
background, there is no particle production by the black hole.
The situation is more like instabilities in the background of a
negative mass Schwarzschild solution in 4-dimensions, as we will
indicate later.

\chapter{ The unstable modes}
We will investigate the question of instability in the
$ \left( SL (2,R) / U(1) \right)_k$ formulation
of the Euclidean black hole background [\WITTEN]. We search for Virasoro
primaries of this CFT which are relevant operators. If we have the
dimensions $ h = \bar{h} < 1$ for a primary,  it corresponds
to a negative eigenvalue off-shell mode, {\it i.e.} a small
deviation from the classical solution which lowers the string field
action. This solution, then, is not a local minimum. Calculation of
one-loop correction around such a stationary point is divergent
and has to be defined by analytic continuation. In that process
the contribution might become imaginary. \foot {This piece is a
volume independent contribution to the one-loop partition function.
The bulk contribution is the same as that of
Liouville coupled to a compact
boson with radius three halves the self-dual radius.}
This indicates that the
classical solution is unstable under quantum fluctuations and
most of the probability would be concentrated around some other
solution which is a true minimum of the action. In other words,
a quantum state peaked around this classical solution will decay
into some other more stable state.

To obtain the Euclidean black hole solution, we look at an $SL
(2,R)$ {\sl WZW}-model with level $k = 9/4$, and gauge an
$SO(2)$ ({\it i.e.} $U(1)$) subgroup. The stress tensor of the
gauged theory is given by
$$
      T(z) = T^{SL(2)} (z) - T^{U(1)} (z)
      \eqn \T
$$
Hence
$$
      L_0 = L^{SL(2)}_0 - L^{U(1)}_0
      \eqn \L
$$
For a chiral primary $|j,m \rangle$ characterized by $SL(2)$
isospin $j$ and $J_0^3$ (generator of the $U(1)$ subgroup)
eigenvalue $m$
$$
   L^{SL(2)}_0 |j,m \rangle   =  -{ {j(j+1)}\over {k-2}}|j,m \rangle
      \eqn \sl
$$
and
$$
   L^{U(1)}_0 |j,m \rangle   =  -{ {m^2}\over {k}}|j,m \rangle
      \eqn \u
$$
The scaling dimension therefore is
$$
   h_{j,m} = -{ {j(j+1)}\over {k-2}} + { {m^2}\over {k}}
        \eqn \h
$$
When $j$ corresponds to the principal continuous series ({\it
i.e.} $ j = - { {1} \over {2} } + i \lambda, \; \lambda \in
R $), $h_{j,m} \ge 1$. So, there are no relevant operators for
these $j$'s. For real $j$ however it is possible to get low
enough scaling dimensions which will make the operator relevant.

In order to decide the question of whether such real $j$'s are
allowed, requires the knowledge of the full operator content. In
absence of this information, we have some indirect
arguments about their presence. First, the modular invariant
combination that can be formed from just the $j$'s belonging to
the prinipal continuum series only , seems to represent the
physics of a flat cylindrical target space where the noncompact
direction has a background charge. From the spacetime point of
view, in the background of the semi-infinite cigar solution
(which is how the Euclidean black hole solution looks) [\WITTEN],
we expect square-integrable
solutions concentrated around the cigar tip, which decaying exponentially
towards the flat cylinder end of the cigar.
Dijkgraaf {\it et al} [\DVV] call
these offshell modes the bound states. Let $r$ be the coordinate
along the length of the semi-infinite cigar and $\theta$ be the
coordinate along the periodic direction. $r$ goes to infinity as one goes
towards the asymptotically flat region. $\theta$ has period $2\pi$.
The asymptotic form of the vertex operator in these coordinates, if chosen
appropriately, is $e^{jr+im\theta}$. Hence the solutions which decay at
large $r$ correspond to real $j$. Once certain real $j$ solutions are
there, it seems possible to generate larger values of $j$ by fusion.

The theory at the flat cylinder end looks like a compact boson
$\theta$ times a free scalar field $r$ with a background
charge. The radius of the compact boson is $\sqrt{k} R_0=3 R_0 /2$
where $R_0$ is the
self dual radius. Taking this asymptotic theory seriously, we have, {\it a
la} Dijkgraaf {\it et al.}[\DVV],
$$
    m = { {p}\over{2} } + {{qk}\over{2}} \qquad p,q \in Z
        \eqn \dijk
$$
For continuous representation, as already stated,
$ j = - { {1} \over {2} } + i \lambda, \; \lambda \in
R $ and for discrete representation, $ j + |m| \in Z$.

If we look at the vertex operators of zero spin, the modes which
will be considered in string field theory, then either $q$ is
zero  or $p$ is zero. The first case corresponds to winding modes and the
second to momentum modes.  For these two kinds of spin zero vertex
operators, the operator $L_0 -1$ written as a differential
operator in the target space coordinates looks like:

$$
(L_0 -1)_{\rm mom}
=-4\left[{{\partial ^2} \over {\partial r^2}} +(\coth ^{2} {{r} \over {2}}
-{{8}\over{9}}) {{\partial ^2} \over {\partial \theta^2}} + {{1} \over {16}}
\coth ^{2} {{r} \over {2}} + {{1} \over {16}} \tanh ^{2} {{r} \over {2}}
-{{1}\over{8}}\right]
\eqn \mom
$$
for the momentum modes and
$$
(L_0 -1)_{\rm wind}
=-4\left[{{\partial ^2} \over {\partial r^2}}+(\tanh ^{2} {{r} \over {2}}
-{{8}\over{9}}){{\partial ^2} \over {\partial \theta^2 }} + {{1} \over {16}}
\tanh ^{2} {{r} \over {2}} + {{1} \over {16}} \coth ^{2} {{r} \over {2}}
-{{1}\over{8}}\right]
\eqn \wind
$$
for the winding modes in the coordinates of the dual space-time
($\theta$ has a different periodicity here).
Note that these differential operators are being applied
on the tachyon wave-function $\tilde T$, where $\tilde T=Te^\Phi$, $T$
being the usual tachyon and $\Phi$ is the dilaton background. If we take
$\tilde T = R_{j,m}(r)e^{im \theta}$ the `radial' operator would be
obtained from the above mentioned by replacing
${{\partial ^2} \over {\partial \theta^2}}$ by $-m^2$. These operators
are one dimensional Schr\"odinger operators on half-line, since $r$
goes from zero to infinity. The potential in the Schr\"odinger
operator, for the winding mode, has an attractive $1/r^2$ . For the
tachyon mode at non-zero $m$, that is not the case. When $m=0$, the
solutions which are smooth at the tip of the cigar, are not
square-integrable. Therefore,it seems that the
 square-integrable modes come only from the winding sector.
  They can have arbitrarily negative values of $L_0-1$.

It is interesting to look at the same problem from the dual space-time.
 The Euclidean black hole solution goes to a solution
with a naked singularity. The cigar becomes a trumpet with an
infinitely large rim [\GIV,\DVV]. Since under duality transformation the
momentum and the winding modes get exchanged, we now have
particle-like modes giving rise to instabilty.

Something very similar happens in the 4-d Schwarzschild solution
with negative mass. The Laplacian operator in this background
has negative modes. It becomes obvious if one goes to the
equation for the radial part of the eigenvalue problem.
By changing to some new variable $r_\ast = r_\ast(r)$, the
problem can be made equivalent to a half line Schr\"odinger
problem with an attractive potential going to $-\infty$ near the
origin.

The solution we considered here is an Euclidean solution with no
conical singularity. It is tempting to generalize to a solution
which has a conical singularity. This allows us to take the
radius of the asymptotic compact boson $R = 3 R_0/2\xi$, where
$\xi$ is an arbitrary positive number. Allowed $m$ values would become
$$
    m = { {p\xi}\over{2} }  + { {qk}\over{2\xi} }
        \eqn \conical
$$
Most of the qualitative physics should not change. It appears
that for $\xi < 1/2$, {\it i.e.} $R > R_0$, momentum mode
instabilities are also possible if one follows the arguments
involving the potential in the related `radial' operator.

So far we have talked of instabilities only in chiral primaries.
These are the tachyon modes. To get the unstable modes of higher tensor
fields which lead to instability we look for some current algebra
secondaries which are Virasoro primaries. Nelson and Distler found an
interesting isomorphism between different discrete representations which
keeps the dimension unchanged but changes the oscillator number
[\DN]. In the
free field representation, the stress tensor is
$$ T(z) = -{ {1}\over {2}
} \partial \sigma \partial \sigma + { {1}\over {2} } a \partial^2 \sigma +
{ {1}\over {2} } \partial \phi \partial \phi \eqn  \free $$
$\phi$ is a
compact boson and $a = \sqrt{ { {2} \over {k -2} } }$.  The chiral
primaries are
$$ V_{j,m} = e^{ - j a \sigma + i m \alpha \phi} \eqn \Vjm
$$
$\alpha = \sqrt{2/k}$. Note that $r \sim -a\sigma$ and $\theta
\sim \alpha \phi$ in the asymptotically flat region.
For the so-called $D^+$ representations, $j - m
< 0$. Let $S^-$ be the operator $ \oint dz~ e^{\sigma(z)/a - i
\phi(z)/\alpha} $.
$$\eqalign{
 S^- V_{j,m}(0) =& \oint dz~ z^{j-m} : V_{j,m}(0)
e^{\sigma(z)/a - i \phi(z)/\alpha}:    \cr
\sim & \left[ (J^+)^{m-j+1} +
{\rm other~ terms} \right] e^{-a (j - 1/a^2)\sigma + i \alpha (j
-1/\alpha^2) \phi} \cr}
\eqn \S $$
where $J^+ \sim \partial \sigma e^{i \alpha \phi} $.
These are higher level states in the
representation they call $\widetilde{D}^-$.

If these operators are allowed in the spectrum of the conformal
field theory, then they correspond to generically offshell modes
of higher tensor fields. Since the $L_0$ eigenvalue is
unchanged, if we start with a relevant tachyon operator, we end
up with a relevant operator from the higher level.

\chapter{ Conclusion}
We indicate the operators in the conformal field theory which
correspond to unstable modes in the target space field theory.
It is interesting to compare this instability with the
Kosterlitz-Thouless ( KT) instability [\KT].

For 26 dimensional bosonic string having one compact direction
(one can think of this as a finite temperature theory), the
vertex operator of the form $e^{ip\cdot x} \times$ (winding mode
of winding number $n$) has dimension
$$
     h = { {p^2}\over {2} } +
           \left( { {nR}\over {2R_0} } \right)^2
$$
For $ p \to 0$ and $n = 1$, this operator becomes relevant for
$R < 2 R_0$. Thus there is KT transition above a certain
temperature.

In the case at hand, the situation is more tricky. Here,
$$
     h = - 4 j (j + 1) +
           \left( { {nR}\over {2R_0} } \right)^2
       = 1 + \left( { {nR}\over {2R_0} } \right)^2  -(2 j + 1)^2
$$
If $ j = - { {1} \over {2} } + i \lambda$, no matter how small
$R$ is, the operator is irrelevant. So, the instability is a
consequence of $j$ being real for discrete representations.

Let us believe that we can understand the question of metastability
of Minkowsky gravity by doing a semiclassical analysis of the
corresponding Euclidean path integral [\GPY].
We would like to speculate that the string field theory
path integral around the
Euclidean black hole  or its dual is dominated by the contribution
from the region around the corresponding flat infinite cylinder solution
with the same radius for the asymptotic compact boson. Presumably
this space-time has a tachyon background also. This will indicate that
the Minkowski black hole ( or a naked singularity, which is dual to
the region outside the horizon) decays into flat finite temperature
space, the temperature being the same as the Hawking temperature
of the curved space
\foot {This possibility was indicated by S. Wadia in a talk given
at MSRI, Berkeley, earlier this year.}.
Presumably, for the flat infinite cylinder, the real $j$ vertex
operator solutions are not allowed, since they diverge at one of
the ends. Exclusion of these operators would make this solution
stable. It would be interesting to understand the relation between
this viewpoint with the discussion of matrix model on a circle by
Gross and Klebanov [\GKCIRC].

{\bf Acknowledgements}: I am extremely grateful to S. Das, G. Mandal and
A. Sen for detailed comments on various aspects of this work. I thank A.
Dhar, D. Jatkar, S. Mukhi and T. Padmanabhan for various discussions. It is
a pleasure to thank S. Wadia for collaboration in an early phase of this
project and for his continuous encouragement. I am
also indebted to S. Mukherji and D. Choudhury for the support and help
they provided.

{\bf Note added}: After this work was finished, we received a preprint
by J. Ellis, N. Mavromatos and D. Nanopoulos, CERN Preprint,
CERN-TH.6309/91, ACT-53, CTP-TAMU-90/91/. However, they seem to be
referring to a different source of instability.

\refout

\end